\documentclass[sigconf]{acmart}

\usepackage{booktabs} 

\setcopyright{rightsretained}

\acmConference[ICTIR 2018]{ACM SIGIR}{Sep. 2018}{Tianjin, China}
\acmYear{2018}
\copyrightyear{2018}

\acmArticle{8}

\editor{Jennifer B. Sartor}
\editor{Theo D'Hondt}
\editor{Wolfgang De Meuter}

\begin{document}
\title{Exploiting "Quantum-like Interference" in Decision Fusion for Ranking Multimodal Documents}

\author{Dimitris Gkoumas}
\affiliation{%
  \institution{The Open University}
  \city{Milton Keynes}
  \country{UK}}
\email{dimitris.gkoumas@open.ac.uk}

\author{Dawei Song}
\affiliation{%
  \institution{The Open University}
  \city{Milton Keynes}
  \country{UK}}
\email{dawei.sogn@open.ac.uk}

\begin{abstract}
Fusing and ranking multimodal information remains always a challenging task. A robust decision-level fusion method should not only be dynamically adaptive for assigning weights to each representation but also incorporate inter-relationships among different modalities. In this paper, we propose a quantum-inspired model for fusing and ranking visual and textual information accounting for the dependency between the aforementioned modalities.  At first, we calculate the text-based and image-based similarity individually. Two different approaches have been applied for computing each unimodal similarity. The first one makes use of the bag-of-words model. For the second one, a pre-trained VGG19 model on ImageNet has been used for calculating the image similarity, while a query expansion approach has been applied to the text-based query for improving the retrieval performance. Afterward, the local similarity scores fit the proposed quantum-inspired model. The inter-dependency between the two modalities is captured implicitly through "quantum interference". Finally, the documents are ranked based on the proposed similarity measurement. We test our approach on ImageCLEF2007photo data collection and show the effectiveness of the proposed approach. A series of interesting findings are discussed, which would provide theoretical and empirical foundations for future development of this direction. 
\end{abstract}

%
%



\maketitle

\section{Introduction}

Fusion and assessment of relevant multimodal information are open research issues leading to many challenges. One of the most critical problems is the semantic gap between low-level features and high-level user's information need. Specifically, the semantic gap arises from the fact that computers and human understand the content in different ways. Besides the semantic gap, multimodal fusion is another crucial task. Because of heterogeneity of multimodal information, composite systems fail to capture inter-relationships among different modalities. Additionally, many systems use a standard metric system to measure and evaluate the similarity among different modalities, although humans use completely different models to assess multimedia data similarity \cite{farag2003human}.

Researchers have applied various techniques, such as relevance feedback, query formulation, statistical approaches and machine learning, in many cases taking into account human behaviour for bridging the semantic gap \cite{atrey2010multimodal,khokher2011content,bokhari2013multimodal}. Considering that low-level features do not directly express user's high-level perception, the query formulation process in a multi-modal information retrieval (IR) system is a difficult task \cite{urban2006adaptive}. On the other hand, relevance feedback is a more efficient way for bridging the semantic gap in a multimodal system learning a query through user interactions.

Multimodal fusion is another way to alleviate the semantic gap. Currently, the state-of-the-art multimodal fusions are \textit{model-based} approaches that explicitly address fusion in their construction - such as kernel-based approaches, graphical models, and neural networks. Deep learning, which is a subfield of machine learning, tackles the problem of the semantic gap at a feature level by learning joint embeddings among multiple modalities \cite{salvador2017learning,baltruvsaitis2018multimodal}. This approach is very common to the cross-modal information systems where each modality is mapped to a common subspace \cite{baltruvsaitis2018multimodal,rasiwasia2010new}. Then, for instance, a set of images can be retrieved in response to a text query and vice versa. In a similar way, tensor decomposition approaches such as the Tucker decomposition \cite{ben2017mutan} and tensor factorization \cite{arora2016cross} have been used for capturing inter-dependencies between visual and textual features. Both deep learning and tensor decomposition approaches fuse different modalities at a feature level. The feature level fusion, known as early fusion, integrates features after extraction. The early fusion is advantageous in that can exploit inter-relationships between low-level features of each modality, but it suffers from the curse of dimensionality.

Besides the model-based multimodal fusion, the vast majority has been done using model-agnostic approaches which fall into \textit{early}, \textit{late}, and \textit{hybrid} fusion \cite{atrey2010multimodal}. In contrast to the feature-level early fusion, late fusion at the decision level is more scalable and allows us to exploit the most suitable methods for analyzing every single modality. Additionally, a rigorous late fusion can provide complementary information and increase the accuracy of the  overall decision. However, the main disadvantage of late fusion is that it fails to measure the mutual information among different modalities. In this paper, we argue that "quantum interference" does model and cater for the dependencies among different modalities at a decision level fusion. Finally, hybrid fusion attempts to exploit the advantages of both the described methods in a common framework combining outputs from early fusion and individual unimodal predictors. However, it requires the training of many models, making the training pipeline more complicated compared to late and early fusion.

Quantum-inspired models with an interference term have been used not only for retrieving information but also for explaining some non-classical phenomena. The mathematical framework of quantum theory has been much more successful to model "fallacies of human reasoning" in cognitive and social sciences. A whole set of experimental findings revealed that set-theoretic structures, like those used in classical (fuzzy) logic and probability theory, are not capable of explaining non-classical phenomena \cite{townsend2000exploring,osherson1981adequacy}. A whole set of experiments revealed that people in some cases tend to overestimate the joint probability (conjunction) or underestimate the probability of a union of events (disjunction) violating the Kolmogorovian probability. One of that most well-known experiments is the Tversky and Kahneman's experiment \cite{tversky1983extensional} discovering through Linda's story the "conjunction fallacy" violating the monotonicity law of probability. Specifically, they found that the probability of conjunction $P(A \cap B)$ is higher than the probability of its constituent P(B), while the probability of disjunction $P(A \cup B)$ is lower than the probability of its constituent P(A). Decision making is a constructive process involving a transition among judgments disturbing each other, introducing uncertainty and order effects. Thus, quantum probability, as a generalized framework of the classical probability, can interpret through quantum interference term more successfully decision-making processes. Specifically, any interference among different judgments comes from the combination of individual judgments, that is, interference exists because a user is in an ambiguous superposition state before the final decision.  

Except for cognitive cognition, quantum-like models with an interference term have been explored in information retrieval (IR), as well. Zuccon and Azzopardi proposed an IR model based on quantum probability ranking principle for subtopic retrieval, where novelty and diversity are required \cite{zuccon2010using}. The dependencies between documents were captured through a "quantum interference" term. However, the interference term was approximated through a similarity function between documents. In an empirical evaluation, they found that the proposed model outperformed models based on classical probability ranking principle. Wang et al. investigated the phenomenon of quantum-like interference in document relevance judgment via an extensive user study \cite{wang2016exploration}. The existence of quantum-like interference was verified through the law of total probability and the order effect. In \cite{xu2008order}, the order effect of relevance judgement was referred to as different relevance perceptions of a document when it appears in different positions in a list. The experimental results revealed the existence of the quantum-like interference. Moreover, Zhang et al. proposed a multimodal decision fusion strategy inspired by quantum interference for multimodal sentiment analysis \cite{Zhang2018}. The proposed model significantly outperformed a wide range of state-of-arts methods. However, the interference term was treated as a single parameter and adjusted experimentally.

In a similar way, we argue that the quantum probability theory \cite{zuccon2009quantum}, which naturally includes interference effects among events, can capture the dependencies among different modalities for fusing multimodal information at a decision level.  

The remainder of the paper follows. In Section II we mention the motivation of the investigation, while in Section III we explain some basic concepts of quantum theory and then formalize the proposed model. Section IV reports all the experiment settings. In Section V we report the results, while in Section VI we present an analysis of the results. Finally, Section VII concludes the paper and suggests some future research directions.

\section{MOTIVATION }
The fusion of multimodal data, which is the problem of making decisions, possesses many challenges. In such a case, a user judges multimodal information based on his cognition of content through concurrent multiple channels corresponding to different modalities. For instance, let's assume a simple case where a multimodal document consists of visual and textual information. In such a case, the relevance of a multimodal document for a given query is based on the visual and textual content. Let's also denote p(R|T) the probability for a multimodal document $D_M$ to be relevant to a multimodal query $Q_M$ concerning the textual information, and p(R|V) the corresponding probability concerning the visual information. The simplest and most used method for fusing information is the linear weight fusion using sum operators \cite{atrey2010multimodal}. The issue of finding the optimal weight for different modalities is an open research issue. To dumb the problem down, we assume that both the visual and textual channels are equally important. Then the total probability of relevance $p_T(R)$ summing up to 1 is computed by the well-known formula: 
\begin{equation}
p_T(R) = 0.5 \cdot p(R|V) + 0.5 \cdot p(R|T)
\end{equation}
On the other side, J. Busemeyer et al. \cite{busemeyer2009empirical} expose two different approaches for building up probabilistic dynamic systems: The first one is based on (a) Markov Theory and the second one on (b) Quantum Theory. The main difference between the Markov and Quantum theory is that the latter does not obey the law of total probability. They found out that there are cases that quantum probability theory could better model and interpret irrational phenomena. 

Being inspired by the aforementioned work, we believe that the dependencies between the visual and textual modality at a decision level fusion could be modelled by the quantum probability theory \cite{zuccon2009quantum}. The potential difference between the classical and quantum probabilities can be explained as interference between the visual and textual channels. If there really exists a "quantum-like interference" phenomenon between the visual and textual channels, then the quantum-inspired model should rank the multimodal document with a higher precision. To the best of our knowledge, this is the first time that a decision-level fusion tries to capture implicitly the dependencies among different modalities in the multimodal information retrieval task.

\section{QUANTUM-LIKE MULTIMODAL DECISION FUSION}

Our basic assumption is that at first the user, before deciding about the relevance of a document, is being in a superposition state between the visual and textual information. Because of the superposition state, the judgement of document introduces "quantum-like interference" between the visual and textual channels.

To elucidate the idea of "quantum-like interference" for decision-making tasks, we introduce Townsend's paradigm for studying the interactions between categorization and decision making \cite{busemeyer2012quantum}, which fits well for testing "quantum-like models". As a matter of fact, Townsend's paradigm assumes that the enrolled identities (in this case the textual information, visual information, and relevance of document) can be expressed as transition probabilities.

\begin{figure}[h]
\centering
\includegraphics[scale=0.25]{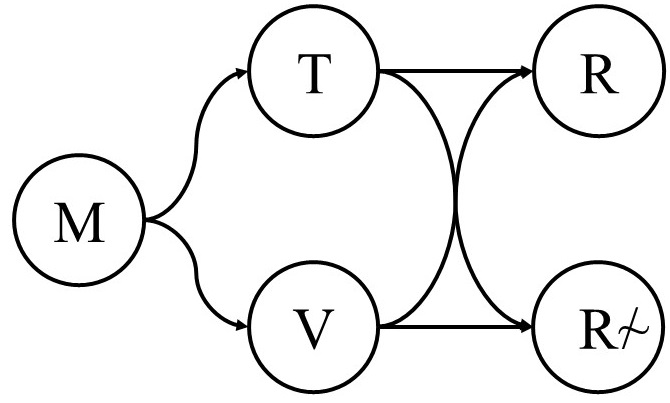}
\caption{Diagram representing category - decision-making task. Starting from the multimodal document, the user interacts with the textual or visual information.}
\end{figure}

Figure 1 illustrates the basic idea tailored to a multimodal information retrieval system, which is indeed an analogy of the well-known double-slit experiment \cite{feynman1965lectures}. At the very beginning, the user is being into an uncertain state M representing a multimodal document. From this statement, the user can transit from M to the T state representing the textual information of the document, with a probability p(T|M). S/he could also transit from M to the V state representing the visual information of the document, with a probability p(V|M). From the state T, the user can transit to the R state corresponding to the relevance of the multimodal document with a probability p(R|T). Similarly, s/he could also transit from the V to R, with a probability p(R|V). Alike transitions can be observed from the T and V states to the R$\nsim$ state representing the non-relevant state of the multimodal document. In our case, the latter transition is entirely dismissed and we focus only on the transitions ending up in the relevant state (Figure 2).

\begin{figure}[h]
\centering
\includegraphics[scale=0.25]{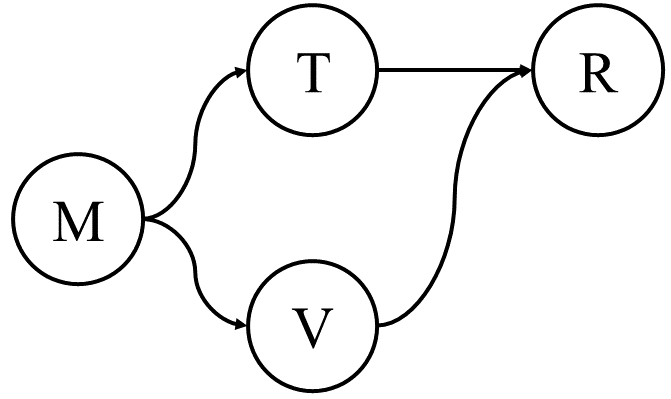}
\caption{Ranking only relevant documents.}
\end{figure}

Feynman's rules can mathematically model the above transition from the M to R state (Figure 2). According to the rules, if we do not know which path has crossed from the M to R state, then we firstly sum up the path of amplitudes across all possible paths and afterwards, we take the squared magnitude of the sum (Formula 2).  In formula 2, the third term is called interference term, while $\cos{\theta}$ is the difference in phase between the $M \rightarrow T \rightarrow R$ and $M \rightarrow V \rightarrow R$ path.

\begin{align}
p(M \rightarrow R)  &= |\langle T|M \rangle \cdot \langle R|T \rangle + \langle V|M \rangle \cdot \langle R|V \rangle|^2 \nonumber \\ \nonumber   &=  |\langle T|M \rangle|^2 \cdot |\langle R|T \rangle|^2 + |\langle V|M \rangle|^2 \cdot |\langle R|V \rangle|^2 \\ &+ 2|\langle T|M \rangle \langle R|T \rangle \langle V|M \rangle \langle R|V \rangle| \cdot \cos{\theta}
\end{align}

The probability associated to the events of passing through the textual channel only and being judged as a relevant multimodal document is equal to the squared magnitude of the amplitude for the transition $M \rightarrow T \rightarrow R$. Similarly, the probability when a user judges the relevance of a document depended on visual content only is equal to the squared magnitude of the amplitude for the transition $M \rightarrow V \rightarrow R$. Then, involving probabilities, Formula 2 results in

\begin{align}
    p(M|R) &= p(T|M) \cdot p(R|T) + p(V|M) \cdot p(R|V)  \\ \nonumber     
            &+ 2 \sqrt{(p(T|M)p(R|T)p(V|M)p(R|V)}  \cdot \cos{\theta}
\end{align}

In a multimodal information retrieval task, p(R|T) stands for the similarity of textual information, and p(R|V) the visual similarity respectively. The probabilities p(T|M) and p(V|M) are weighted parameters for declaring the importance of visual and textual information over the judgment process correspondingly. Given that the proposed model fits with real numbers, $\cos{\theta}$ can range from -1 to 0 to +1, producing negative, neutral, or positive interference each in order. In our case, when interference term results in a positive value implies a constructive dependency between the visual and textual information reinforcing the ranking score. In the same way, when interference term results in a negative value indicates a cancellation effect between the two channels undermining the ranking score, that is, the visual and textual information conflicts. Finally, when interference term is equal to 0 means that the visual and textual channels do not share any common information, i.e., they are mutually orthogonal, eradicating any interference effect. In that aspect, the proposed model reduces to the classical model computing the ranking score based on the law of total probability (Formula (4)).

\begin{equation}
p(M|R) = p(T|M) \cdot p(R|T) + p(V|M) \cdot p(R|V)
\end{equation}

For adapting the interference dynamically between the visual and textual channels, we define an upper $(T_U)$ and a lower $(T_L)$ probability thresholds. For the sake of simplicity, in formula 3, we denote the joint probability $p(T|M) \cdot p(R|T)$ as p(T) and $p(V|M) \cdot p(R|V)$ as p(V), respectively. In total, there are 4 possible rules for investigating any constructive interference or cancellation effect between the visual and textual modality (Figure 3), and are defined as follows:

\begin{figure}[h]
\centering
\includegraphics[scale=0.3]{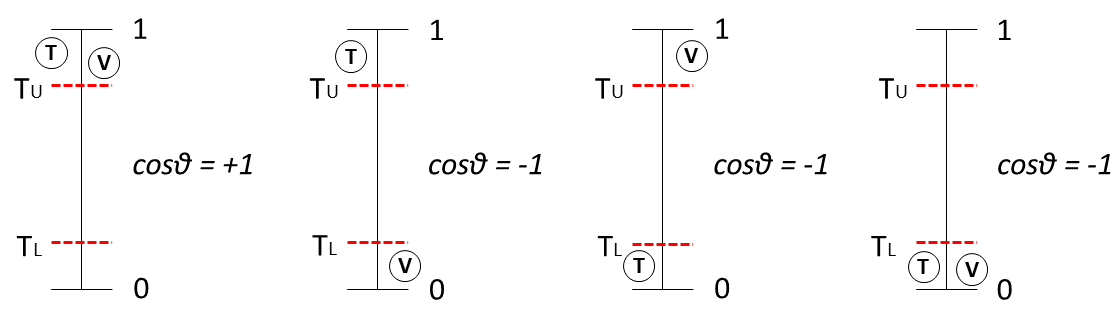}
\caption{Rules for calculating constructive and deconstructive "quantum-like interference" between the visual and textual channel.}
\end{figure}

\begin{itemize}
  \item IF |p(T) > $T_U$|  AND |p(V) > $T_L$|  THEN $\cos{\theta}$ = +1, that is, constructive interference.
  \item IF |p(T) > $T_U$|  AND |p(V) < $T_L$|  THEN $\cos{\theta}$ = -1, that is, deconstructive interference.
  \item IF |p(T) < $T_L$|  AND |p(V) > $T_U$|  THEN $\cos{\theta}$ = -1, that is, deconstructive interference.
  \item IF |p(T) < $T_U$|  AND |p(V) < $T_L$|  THEN $\cos{\theta}$ = -1, that is, deconstructive interference.
\end{itemize}

In any other case, there is no any "quantum interference" effect.

\section{EXPERIMENT SETTINGS}
\subsection{Dateset}
The proposed model is tested on the ImageCLEF2007 data collection \cite{grubinger2007overview}, the purpose of which is to investigate the effectiveness of combining image and text for retrieval tasks. Out of 60 test queries we picked up randomly 30 ones, together with the ground truth data. Each query describing user information need consists of 3 sample images and a text description. For every query, we created a subset of 300 relevant and irrelevant documents. The number of relevant documents ranges from 11 to 98 ones.

\subsection{Image and Text Representations- Mono-modal Baselines}
Our proposed quantum-inspired decision fusion model is based on the mono-modal retrieval scores. Two mono-modal baselines are used in the experiment.

\textbf{A. Bag-of-words (BoW) model}
For both modalities, the basic representation is a BoW. For images, a bag of SIFT descriptors \cite{sivic2003video} is first extracted from training images, and a visual word codebook is learned through K-means clustering. Then, each image was represented by a vector counting the frequency of each visual word. Text words firstly processed with the NLTK tool \footnote{http://www.nltk.org/} by removing stop words and stemming the text. Then, text converted to TF-IDF vectors with the Scikit-Learn tool \footnote{http://scikit-learn.org/stable/index.html}. The L2-norm was used to normalize term vectors for both image and text vectors. Additionally, for text vectors, we smoothed IDF weights by adding one to document frequencies as if an extra document was seen containing every term in the collection exactly once.  For both modalities, the Cosine function is employed to compute the similarity as the normalized dot product of query and document vectors.

\textbf{B. Feature extraction using enhanced models}
For the visual information, the VGG16 model \cite{simonyan2014very}, with weights pre-trained on ImageNet, is  used, resulting in a feature vector of 2048 floating values for each image. After feature vector extractions, we compute the similarity between the submitted visual query and images in the dataset based on Cosine function. For textual information, a query expansion approach has been applied extending the query with the ten most frequent terms according to the ground truth text-based documents. This indeed corresponds to a simulated explicit relevance feedback scenario. Then, the TF-IDF representation is used for calculating the text-based similarity between the query and documents. 

\subsection{Experiment Settings}
The assignment of different weights to different modalities is always an important and difficult task. However, in the case of the BoW model, for the sake of simplicity, we assume that both visual and textual information are equally important, that is, the p(T|M) and p(V|M) probability is equal to 0.5. Though, in the second approach given that the image-based similarity is five times higher compared to the text-based similarity, we define the probability p(T|M) equal to .2 and the probability p(V|M) equal to .8.

Regarding the lower and upper probability thresholds $T_L$ and $T_U$, in the case of BoW models, we define the lower threshold constant and equal to .01, and the upper threshold equal to the Cosine similarity regarding the text information. That is due to the fact that the range of the text cosine similarity is too narrow compared to the image cosine similarity. In the case of enhanced representation, we define the lower threshold constant and equal to .001.

\subsection{Evaluation Measures}
We use a range of evaluation metrics for measuring the precision and quality of ranking \cite{larson2010introduction}. Concerning the query-specific precision, we adopt the overall precision at all retrieved documents, as well as the precision at 20 and 100 top retrieved documents. We also measure the average precision over all recall points. Moreover, mean average precision (MAP) is used to grasp the difference in overall performance between the baseline and proposed model. Finally, for evaluating the quality of ranking, we make use of the normalized discounted cumulative gain (NDCG) at 100 top retrieved documents. 

\section{Results}
The experimental results are shown in Figures 4-11.

\subsection{Results for the BoW representation}
For the BoW representation, the Quantum-liked model falls behind the baseline one resulting in a decrease of 2.4$\%$ in the MAP. Specifically, the MAP of the quantum model is .25 (SD = .17) while the MAP of the baseline model is .26 (SD = .17). However, there is no statistical difference between the two models (p > .05). There are a few queries where the quantum-like model outperforms the baseline, concerning the precision at the first 100 retrieved documents (Figure 4). However, the model struggles a lot concerning the precision at the first 20 documents, since there is an outstanding number of queries where the baseline model performs much better (Figure 5). The average precision at the first 20 documents is .23 (SD = .23) for the baseline model, and .18 (SD = .2) for the quantum-like model. 

\begin{figure}[h]
\centering
\includegraphics[scale=0.25]{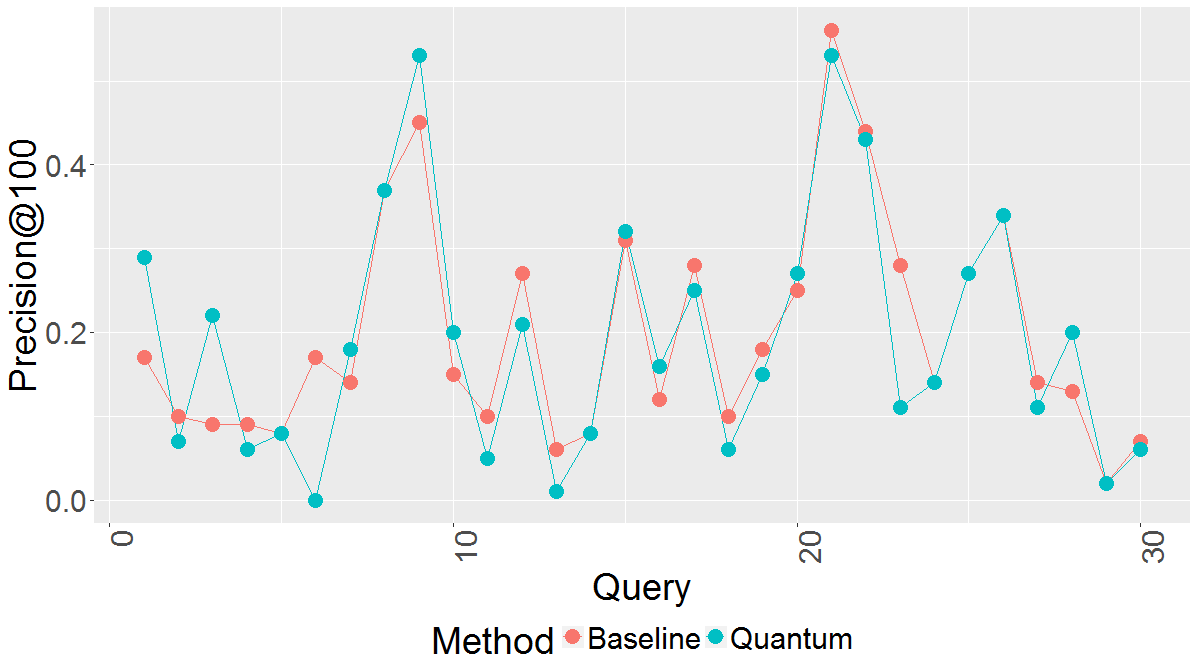}
\caption{Precision at the first 100 multimodal documents for each query respectively -  in case of BoW representation.}
\end{figure}

\begin{figure}[h]
\centering
\includegraphics[scale=0.25]{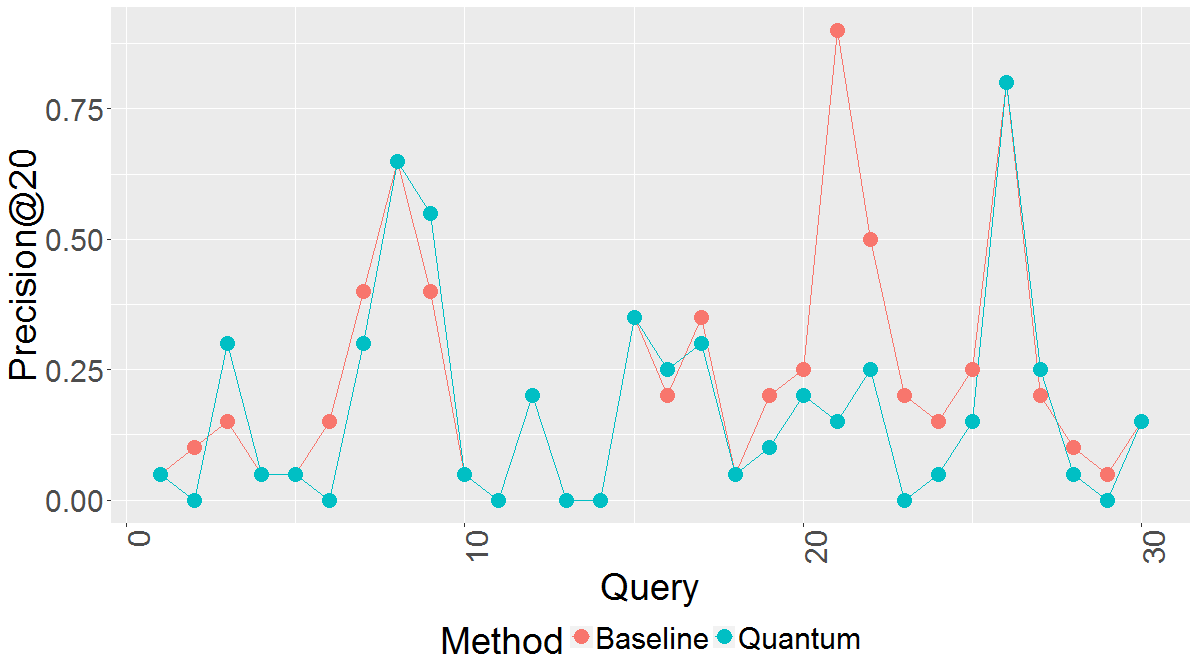}
\caption{Precision at the first 20 multimodal documents for each query respectively -  in case of BoW representation.}
\end{figure}

On the other hand, the quantum-like model is more robust regarding the overall precision.  The mean overall precision is .2 (SD = .1) for the quantum-like model and .18 (SD = .09) for the baseline model. For queries where the quantum-like model does not perform better, it achieves at least a similar performance to the baseline (Figure 6). 

Similarly, the quality of ranking for the quantum-like model is better for some queries (Figure 7). However, the number of queries that the baseline model yields a better quality of ranking at the first 100 documents is higher. The mean NDCG at the first 100 documents is .34 (SD = .25) for the quantum-like model and .37 (SD = .23) for the baseline. Overall, the quantum-like model is neither statistically better nor worse compared with the baseline model (p > .05).

\begin{figure}[h]
\centering
\includegraphics[scale=0.25]{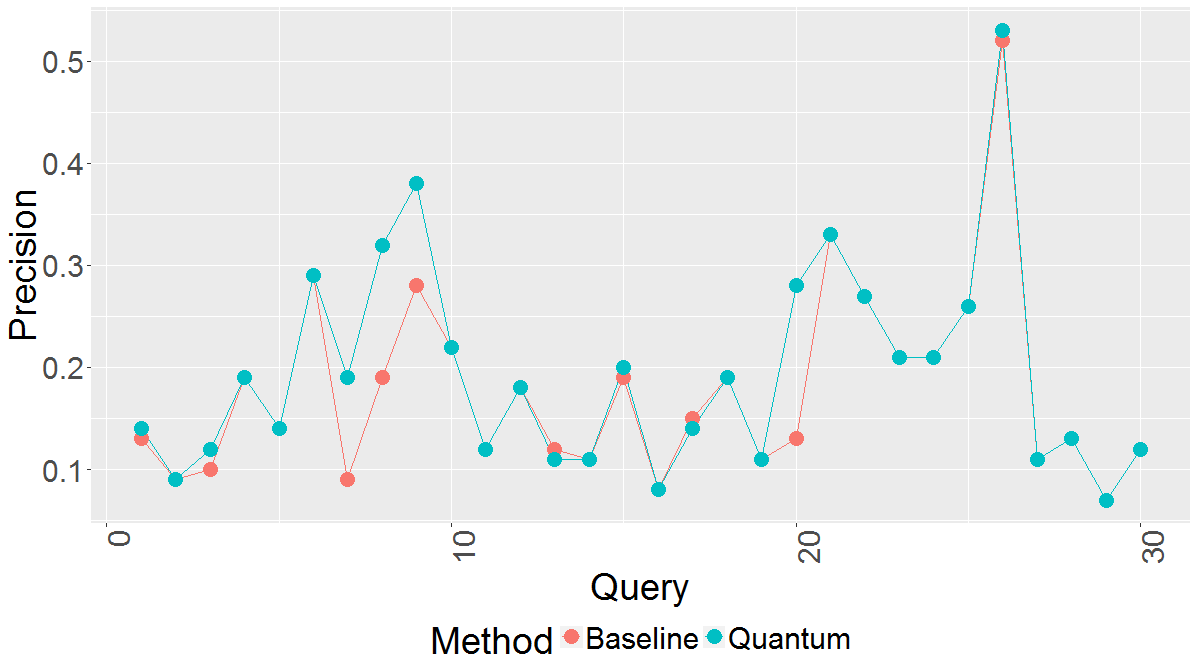}
\caption{Overall precision for each query respectively -  in case of BoW representation.}

\centering
\includegraphics[scale=0.25]{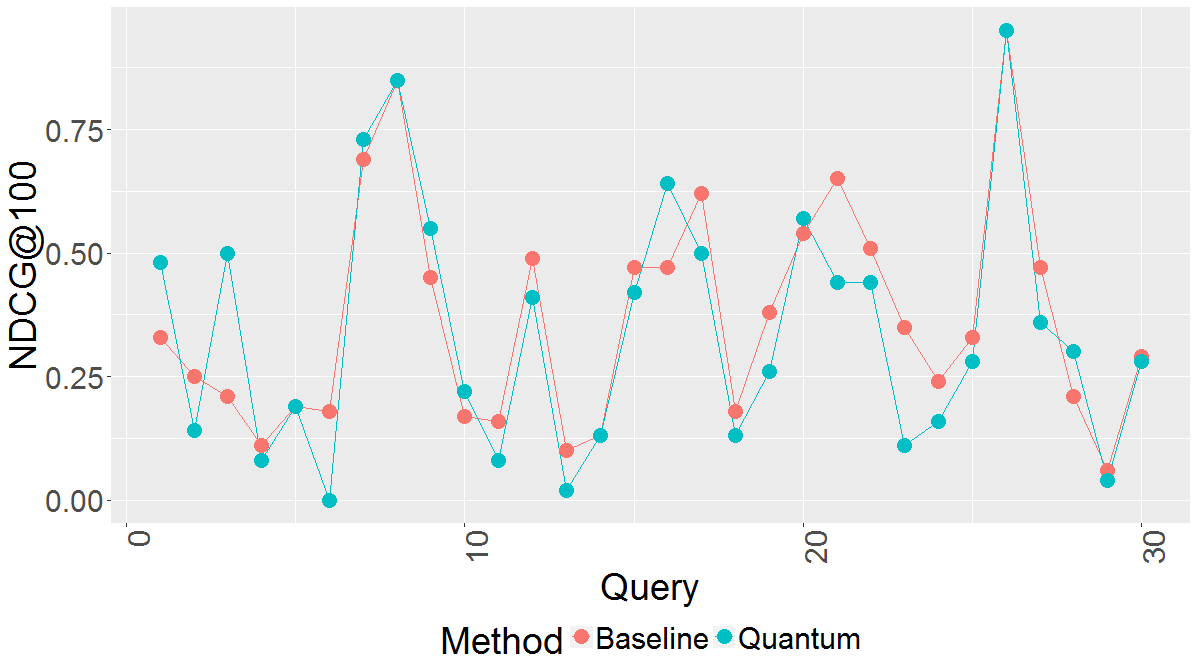}
\caption{NDCG for the first 100 documents for each query respectively -  in case of BoW representation.}
\end{figure}

\subsection{Results for the enhanced representation}
Concerning the second approach which makes use of a pre-trained CNN for the visual representation and a query expansion for calculating the text-based similarity, the Quantum-liked model outperforms the baseline, resulting in a statistically significant increase of 20.3$\%$ in the MAP (p < .05). The MAP of the quantum model is .65 (SD = .18) while the MAP of the baseline model is .53 (SD = .18). The quantum-like model outperforms the baseline since the performance for the majority of the queries is higher for the former model (Figure 8). The average overall precision is .27 (SD = .15) for the quantum-like model and .23 (SD = .13) for the baseline model. However, there is no any statistical difference between the above models.

\begin{figure}[h]
\centering
\includegraphics[scale=0.25]{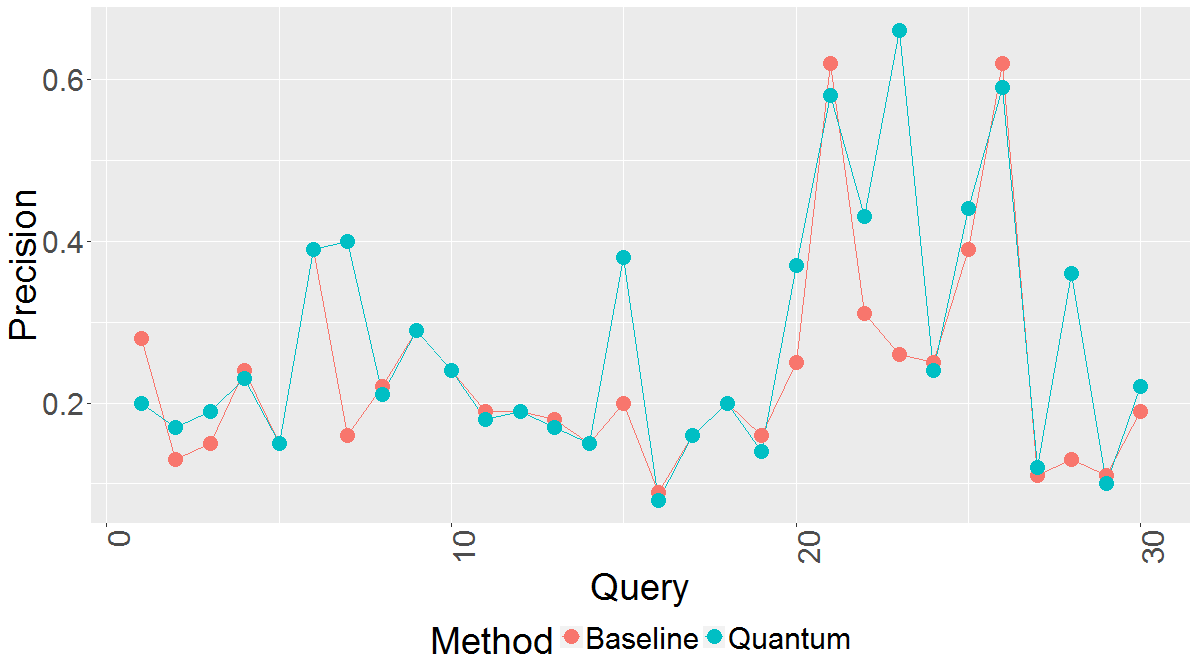}
\caption{Overall precision for each query respectively - in case of enhanced representation.}
\end{figure}

Nevertheless, the quantum-like model performs significantly better than the baseline (p < .05) concerning the precision at the first 20 documents (Figure 9). The average precision at the first 20 documents is .58 (SD = .25) for the baseline model, and .72 (SD = .21) for the quantum-like one. Additionally, the latter model is also more robust regarding the precision at the first 100 documents (Figure 10). The average precision is .36 (SD = .15) for the quantum-like model and .34 (SD = .15) for the baseline model. Also importantly, the quality of ranking at the first 100 documents is higher for the quantum-like model (Figure 11). The mean NDCG at the first 100 documents is .78 (SD = .16) for the quantum-like model and .7 (SD = .15) for the baseline. However, there is no any statistical difference neither in the precision at the first 100 documents nor in the quality of ranking (p > .05).

\begin{figure}[h]
\centering
\includegraphics[scale=0.25]{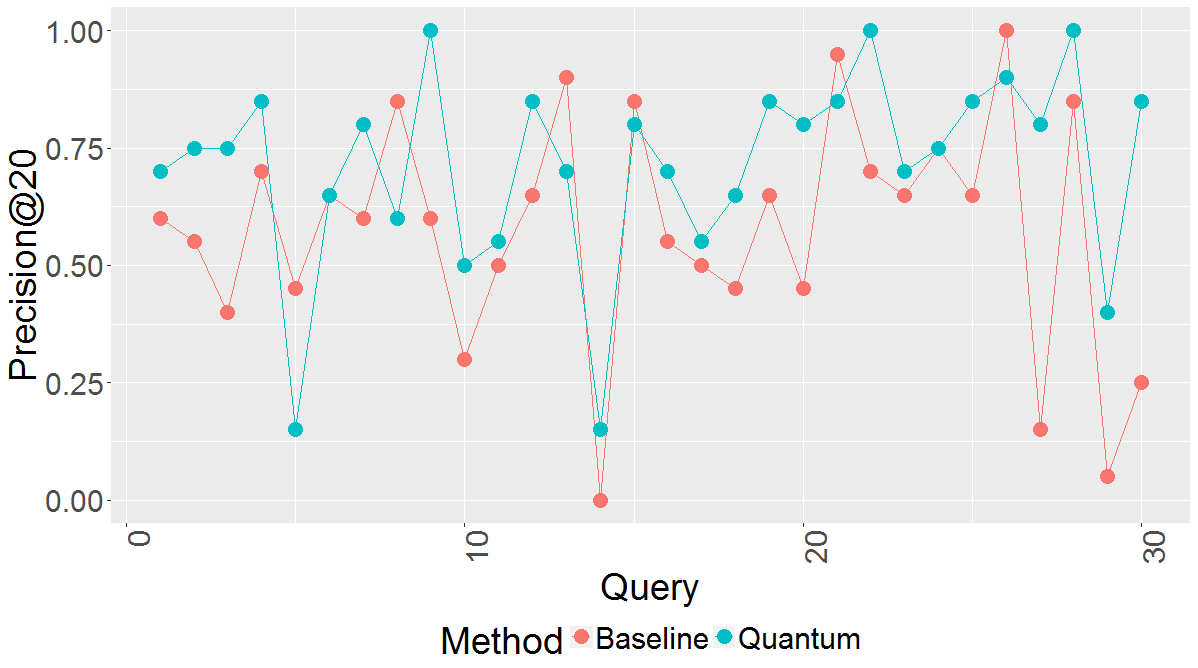}
\caption{Precision at the first 20 multimodal documents for each query respectively - in case of enhanced representation.}
\end{figure}

\begin{figure}[h]
\centering
\includegraphics[scale=0.25]{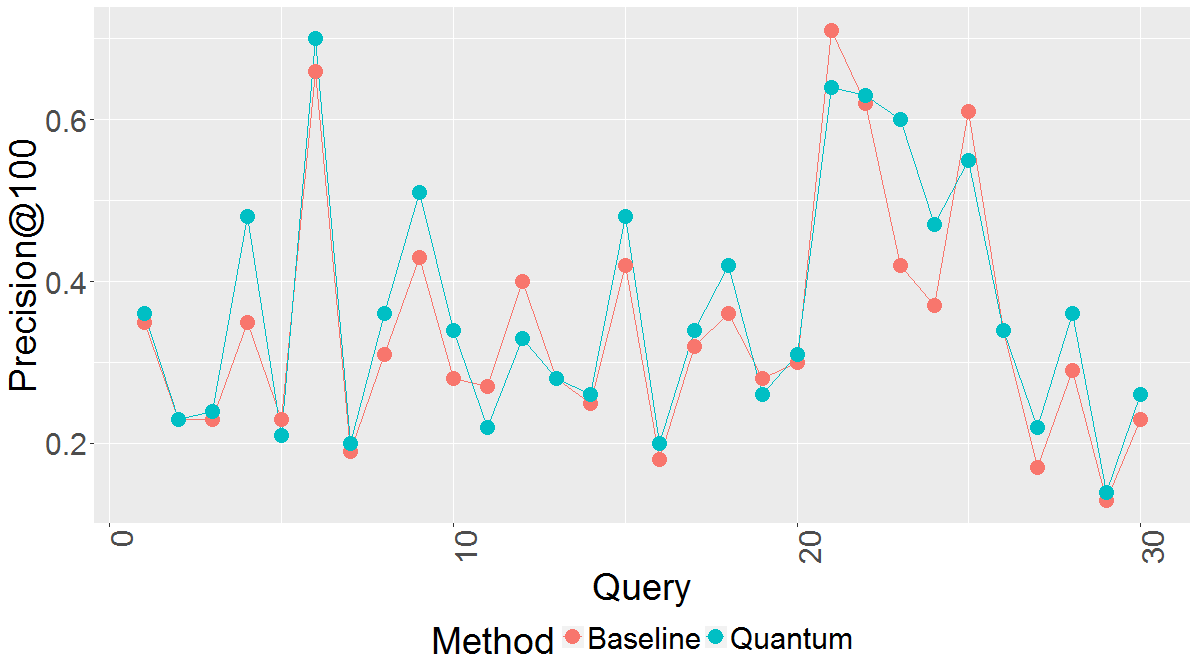}
\caption{Precision at the first 100 multimodal documents for each query respectively - in case of enhanced representation.}
\end{figure}

\begin{figure}[h]
\centering
\includegraphics[scale=0.25]{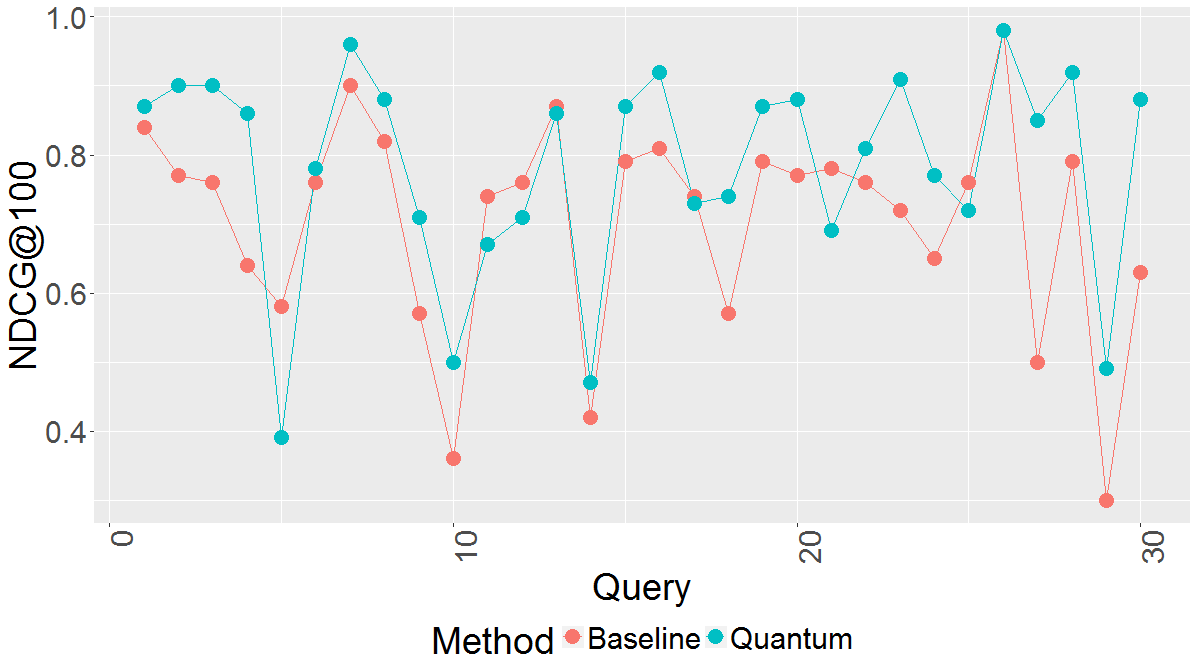}
\caption{NDCG at the first 100 multimodal documents for each query respectively - in case of enhanced representation.}
\end{figure}

In general, for the case of enhanced representation, the quantum-liked model outperforms the baseline and it is also statistically better concerning the precision at the first 20 multimodal documents and MAP (p < .05). 

\section{DISCUSSION}
The first approach for calculating the similarity in the image and text modality before fusion makes use of the bag-of-words representation. The quantum-like model generally underperforms the baseline one. It outperforms the baseline in term of the overall precision, yet not significantly. 

One possible cause of the result is the fact that the BoW model could give a high similarity score to both images and texts that are irrelevant according to the ground truth data. This means that the quantum-like model enhances the ranking scores after the fusion for such "false positive" multimodal documents. Furthermore, ImageCLEF2007 is a pretty challenging dataset. For many queries, the baseline model based on the TF-IDF representation fails to retrieve any textual information, since the documents may not share any common term with the query, resulting in a zero probability (or near-zero probability even after smoothing with the collection distribution) concerning the text similarity. Practically, this means that the multimodal fusion actually based only on the visual modality. In such case, it is difficult for the quantum-like model to capture any dependency between the text and visual modalities.

The above observation motivated us to carry out further experiment with an enhanced representation, which applies a query expansion approach for enhancing the calculation of the text-based similarity. Additionally, a pre-trained deep neural network was used for achieving a higher precision concerning the image information retrieval before the fusion. Initially, we kept the equal weights for the visual and textual channels (i.e., 0.5 for each). We found that the baseline model still outperforms the quantum-like one. Given that the image-based similarity score for a document is often five times higher than the text-based similarity, we assigned the weights equal to 0.8 and 0.2 respectively. By changing the weights, the quantum-like model achieves a significantly better performance than the baseline. Nevertheless, it is not clear if the improvement is resulted from the adjusted weight values or from the enhanced mono-modal representations for calculating the individual similarities before fusing them. This would be an important issue for further investigation.

\section{CONCLUSIONS AND FUTURE WORK }
In this paper, we have proposed a quantum-like model for fusing and ranking multimodal (visual and textual) information accounting for the dependency between the relevance decision-making over the two channels. At first, for the basic bag-of-words representation, the proposed model falls behind, yet insignificantly, the baseline in term of MAP. This may be because the BoW representation is not a very robust method for calculating the text-based and image-based similarities individually. Further, by applying more robust methods and assigning more sensible weights to the visual and textual channels, the quantum-like model is performs statistically better than the baseline. Actually, the quantum-like model results in an increase of 20.3$\%$ in the MAP.

In this paper, we focused on a general first round retrieval process using the quantum-like model as relevance decision model for all multimodal documents against each query. Moreover, we extracted only a relatively small amount of multimodal documents for conducting the experiments. It is worth conducting more experiments with larger scale datasets and also looking at the relevance feedback scenario. For instance, we could re-rank the top 1.000 retrieved multimodal documents by taking into account the explicit or pseudo relevant and irrelevant documents. Finally, some deep learning approaches could be used for extracting text-based representations before fitting the proposed model.

\bibliographystyle{ACM-Reference-Format}
\bibliography{sample-bibliography}

\end{document}